\documentstyle[aaspp4,flushrt]{article}

\catcode`\@=11 
\def\@versim#1#2{\vcenter{\offinterlineskip
        \ialign{$\m@th#1\hfil##\hfil$\crcr#2\crcr\sim\crcr } }}
\newcommand{\Ref}{\hangindent=20pt \hangafter=1 \noindent}
\newcommand{\StartRef}{\hyphenpenalty=10000 \raggedright}

\newcommand{\beq}{\begin{equation}}
\newcommand{\eeq}{\end{equation}}
\def\lsim{\mathrel{\mathpalette\@versim<}}
\def\gsim{\mathrel{\mathpalette\@versim>}}

\def\@versim#1#2{\vcenter{\offinterlineskip
        \ialign{$\m@th#1\hfil##\hfil$\crcr#2\crcr\sim\crcr } }}
\catcode`\@=12 
\begin{document}
\title{Hydrodynamic Drag on a Compact Star Orbiting a Supermassive
Black Hole} 
\author{Ramesh Narayan\footnote{rnarayan@cfa.harvard.edu}}
\affil{Harvard-Smithsonian Center for Astrophysics, 60 Garden St.,
Cambridge, MA 02138}
\medskip
\setcounter{footnote}{0}

\begin{abstract}

The proposed Laser Interferometer Space Antenna is expected to detect
gravitational waves from neutron stars and stellar-mass black holes
spiraling into supermassive black holes in distant galactic nuclei.
Analysis of the inspiral events will require careful comparison of the
observed signals with theoretical waveform templates.  The comparison
could be seriously compromised if non-gravitational torques modify the
orbit of the star.  This paper estimates the torque exerted on an
orbiting star as a result of hydrodynamic interactions with an
accretion flow around the supermassive black hole.  It is argued that
the majority of inspiral events will take place in low luminosity
galactic nuclei in which the mass accretion rate is low and the
accretion occurs via an advection-dominated flow.  The hydrodynamic
torque is negligibly small in such systems and will have no effect on
gravitational wave experiments.

\noindent {\em Subject headings:} Accretion, accretion disks --- black
hole physics --- gravitation --- relativity --- galaxies: nuclei ---
gravitational waves
\end{abstract}

\section{Introduction}

When a compact object spirals into a much more massive object via the
emission of gravitational radiation, the waves carry information on
the multipoles of the central mass (Ryan 1995).  Therefore, by
monitoring the gravitational wave signal one could explore the
spacetime geometry outside the central body, as anticipated by
Abramovici et al. (1992).  If the object is a black hole, it should be
possible to verify its black hole nature and measure its spin
parameter ($a/M$) from the ratios of the multipoles.

Practical application of this idea is expected to become feasible in
the next decade or two.  When a neutron star or a stellar-mass black
hole spirals into a supermassive black hole (SMBH) in a galactic
nucleus, the gravitational waves have periods in the range $10-10^4$
s.  Such periods are well-suited for detection with the proposed Laser
Interferometer Space Antenna (LISA, see Danzmann et al. 1996, Bender
et al. 1996, Folkner 1998, Cutler 1998).  LISA could detect the
inspiral of a $10M_\odot$ black hole into a $10^6M_\odot$ SMBH out to
a redshift $\sim1$.  At these distances event rates of up to several
per year are likely (Hils \& Bender 1995, Sigurdsson 1997).

Both the detection and the analysis of the signals will require a very
clean system with no significant perturbation of the orbiting star.
Is this likely?  In an interesting series of papers, Chakrabarti
(1993, 1996) and Molteni, Gerardi \& Chakrabarti (1994) showed that,
under some cirumstances, the hydrodynamic interaction between the
orbiting star and an accretion disk surrounding the SMBH could be so
strong that the inspiral may be halted altogether and the star may be
trapped in a stable orbit.  Chakrabarti's scenario is an extreme one
that involves a very strong hydrodynamic interaction.  It presumably
occurs only rarely.  However, even if the interaction is orders of
magnitude weaker than the level required by Chakrabarti, it could
still pose a serious problem.  The detection of inspiral events
involves matching the observed signal with a theoretical template
covering many hundreds or thousands of wave periods, and even a tiny
perturbation could be fatal.

Motivated by this, we investigate the following simple question: What
is the expected strength of the hydrodynamic interaction in a
``typical'' galactic nucleus, and how significant is the orbital
perturbation due to this interaction?

Before embarking on a quantitative analysis, we make two points.

First, only a small fraction of galaxies harbor {\it active galactic
nuclei}.  Active nuclei, such as quasars, Seyferts, etc. (see Krolik
1999 for a review), are thought to consist of SMBHs accreting mass at
rates close to or exceeding the Eddington rate $\dot M_{\rm Edd}$.
The accretion is believed to proceed via a thin accretion disk
(Shakura \& Sunyaev 1973, Novikov \& Thorne 1973) when $\dot M\lsim
\dot M_{\rm Edd}$, or a slim accretion disk (Abramowicz et al. 1988)
for $\dot M\gsim \dot M_{\rm Edd}$.  Chakrabarti's work is focused on
such high-$\dot M$ SMBHs; indeed, he explicitly considers
super-Eddington accretion in his analysis.  However, the vast majority
of galactic nuclei are much dimmer than typical quasars or Seyferts.
It has become clear in recent years that these ``normal'' nuclei also
contain SMBHs (e.g. Magorrian et al. 1998), but that the accretion
rates are much below $\dot M_{\rm Edd}$, perhaps as low as $\dot
M\sim10^{-2}-10^{-4}\dot M_{\rm Edd}$.  The lower $\dot M$ implies
that there should be less gas in the vicinity of the SMBH and
therefore a weaker frictional drag on an orbiting star.

Second, at the low values of $\dot M$ relevant for normal galactic
nuclei, there is increasing evidence that the gas flows have a
different character from the flows found in bright nuclei; the
accretion does not occur in a thin or slim disk but via an
advection-dominated accretion flow (ADAF, see Narayan \& Yi 1994,
1995a,b, Abramowicz et al. 1995, Ichimaru 1977, Rees et al. 1982; also
Kato, Fukue \& Mineshige 1998 and Narayan, Mahadevan \& Quataert 1998
for reviews).  For a given value of $\dot M$, the gas density in an
ADAF is significantly lower than that in a thin disk.  This further
reduces the hydrodynamic drag on an orbiting star.

In \S2 of this paper we estimate the hydrodynamic torque on a compact
star orbiting inside an ADAF, and in \S3 we compare this with the
torque that results from gravitational wave emission.  We conclude
with a discussion in \S4.

\section{Hydrodynamic Drag on a Compact Star Orbiting Inside an ADAF}

Consider a compact star of mass $M_c=10m_1M_\odot$ orbiting a SMBH of
mass $M_{\rm SMBH}=10^6m_6M_\odot$.  For simplicity, we assume that the
star is in a circular orbit of radius $R=rR_g$, where $R_g=GM_{\rm
SMBH}/c^2$.  We also work within a Newtonian framework.  This is
adequate in view of the various other approximations we make and
considering that we seek only order of magnitude estimates.  The
orbital velocity of the star is given by the Keplerian formula,
$$
v_K={c\over r^{1/2}}. \eqno (1)
$$

Let the mass accretion rate onto the SMBH be $\dot M_{\rm SMBH}$, and let us
express this in Eddington units:
$$ \dot M_{\rm SMBH}=\dot m\dot M_{\rm Edd}, \eqno (2) 
$$ 
$$ 
\dot M_{\rm Edd}={4\pi GM_sm_p\over\eta\sigma_Tc}=1.4\times10^{24}m_6 
~{\rm g\,s^{-1}}, \eqno (3) 
$$ 
where $m_p$ is the proton mass and $\sigma_T$ is the Thomson
cross-section of the electron.  The parameter $\eta$ refers to the
radiative efficiency of the accretion assumed for the purpose of
defining the Eddington rate; we use $\eta$=0.1 for calculating the
numerical value given in the right hand side of equation (3).  Note
that the choice of $\eta$ in the definition of $\dot M_{\rm Edd}$ is
purely conventional. The actual radiative efficiency in any particular
accretion flow could be different from 0.1.

Let us define also the Salpeter time, which is the mass $e$-folding
time of an object that accretes at the Eddington rate:
$$ 
t_S={M_{\rm SMBH}\over\dot M_{\rm Edd}}={\eta\sigma_Tc\over4\pi Gm_p}
=4.5\times10^7 ~{\rm yr}. \eqno (4)
$$ 
Once again, we have set $\eta=0.1$ on the right.

We assume that the accretion flow around the SMBH occurs via an ADAF.
The key characteristic of an ADAF is that the thermal energy released
as the gas falls into the potential well of the central mass is not
radiated (as in a thin disk) but is retained in the gas and advected
down to the center (Narayan \& Yi 1994).  This causes a number of
important effects on the dynamics of the gas.

First, since all the binding energy is retained as thermal energy, the
gas becomes extremely hot and the temperature approaches the virial
limit.  Equivalently, the isothermal sound speed
$c_s\equiv\sqrt{p/\rho}$, where $p$ is the pressure and $\rho$ is the
density, approaches the Keplerian speed $v_K$.  One consequence is
that the gas does not take up a disk-like configuration (as in a thin
accretion disk), but has a quasi-spherical morphology (Narayan \& Yi
1995a).  Another consequence is that the orbital velocity of the gas
is significantly sub-Keplerian; a typical value is $v_\phi\sim v_K/3$.
Both of these effects simplify our analysis considerably since they
imply that the magnitude of the hydrodynamic drag is insensitive to
the orientation of the stellar orbit relative to the angular momentum
vector of the accreting gas.  The radial velocity $v_R$ of the
accreting gas is quite large, roughly $v_R\sim\alpha v_K$, where
$\alpha$ is the standard dimensionless viscosity parameter (Shakura \&
Sunyaev 1973).  For an ADAF, $\alpha$ typically lies in the range 0.3
(Esin, McClintock \& Narayan 1997) to 0.1 (Quataert \& Narayan 1999).
Thus the radial velocity of the gas is a substantial fraction of
$v_K$.

We can now express the density $\rho$ of the accreting gas in terms of
the mass accretion rate: 
$$
\rho={\dot M_{\rm SMBH}\over 4\pi R^2v_R}
={\dot m\over 4\pi\alpha}{M_{\rm SMBH}\over t_S}
{c^3\over(GM_{\rm SMBH})^2} {1\over r^{3/2}}, \eqno (5)
$$
where we have expressed $\dot M_{\rm SMBH}$ in terms of the Eddington
rate and set $v_R=\alpha v_K$.

Ostriker (1999) has estimated the drag force $F_{df}$ on a mass
$M_c$ moving through a uniform gas of density $\rho$ with relative
velocity $v_{rel}$:
$$ 
F_{df}=-4\pi I\left({GM_c\over v_{rel}}\right)^2\rho, \eqno (6)
$$
where the negative sign indicates that the force acts in the opposite
direction to the velocity of the mass.  The coefficient $I$ depends on
the Mach number, ${\cal M}\equiv v_{rel}/c_s$, of the relative motion.
For ${\cal M}\ll1$, Ostriker finds $I\to {\cal M}^3/3$, while for
${\cal M}\gg1$, $I\to\ln(R_{max}/R_{min})$, where $R_{max}$ is the
size of the gaseous system and $R_{min}$ is the effective size of the
mass $M_c$.

In our problem, $v_{rel}\sim c_s\sim v_K$ and the Mach number is of
order unity.  We could therefore use either the subsonic or supersonic
estimate of $I$.  We use the supersonic estimate as it gives a larger
force and thereby provides an upper limit on the hydrodynamic drag.
We set $R_{max}$ equal to the size of the system, namely the local
radius $R$.  The choice of $R_{min}$ is less obvious since the radius
of the star, the natural choice, is inappropriate for a compact star.
We set $R_{min}$ equal to the gravitational capture radius,
$GM_c/v_{rel}^2$, since Ostriker's linear analysis is valid only for
gas streamlines with impact parameters larger than this radius.
Setting $v_{rel}=v_K$, we then find that
$$
I\sim\ln\left({Rv_{rel}^2\over GM_c}\right)=
\ln\left({M_{\rm SMBH}\over M_c}\right) =
12+\ln\left({m_6\over m_1}\right). \eqno (7)
$$
We use $I=10$ in numerical estimates.

Let us define the hydrodynamic drag time scale $t_{hd}$ to be the
$e$-folding time for the specific angular momentum of the orbiting
star.  Thus
$$
t_{hd}\equiv{M_cv_K\over |F_{df}|}={c^3\over4\pi IG\rho GM_c}
\left({v_{rel}\over v_K}\right)^2{1\over r^{3/2}}. \eqno (8)
$$
Setting $v_{rel}\sim v_K$, and substituting for $\rho$ from equation
(5), we find
$$
t_{hd}={\alpha\over I\dot m}{M_{\rm SMBH}\over M_c} t_S
=10^5{\alpha\over I\dot m}{m_6\over m_1}t_S. \eqno
(9)
$$
Putting in numerical values, this gives
$$
t_{hd}\sim4.5\times10^{10}
{m_6\over\dot m m_1} ~{\rm yr}, \eqno (10)
$$
where we have used ``standard'' parameter values: $\alpha=0.1$,
$I=10$, $\eta=0.1$.

ADAFs are present only for low values of $\dot m\lsim\alpha^2\sim
10^{-1}-10^{-2}$ (Narayan \& Yi 1995b, Esin et al. 1997).  The
galactic nuclei we are interested in probably have even lower
accretion rates: $10^{-4}\lsim\dot m\lsim10^{-2}$.  For such values of
$\dot m$, we see that the hydrodynamic drag on an orbiting star is
extremely small.  In fact, two further effects make the drag even
lower than the above estimate:

\noindent
1. At radii $r\lsim10$, which is the region we are interested in for
gravitational wave studies, the radial velocity of the ADAF is not
$\alpha v_K$ but closer to $v_K$ (Narayan, Kato \& Honma 1997, Chen,
Abramowicz \& Lasota 1997).  This causes the density $\rho$ to
decrease and the time scale $t_{hd}$ to increase by a factor
$\sim1/\alpha\sim10$.

\noindent
2. Blandford \& Begelman (1999) suggested, following earlier work by
Narayan \& Yi (1994, 1995a), that there could be significant mass
outflows from ADAFs.  As a result, close to the SMBH, $\dot m$ may be
one or two orders of magnitude lower than at large radii.  This would
again reduce the hydrodynamic drag on a star orbiting close to the
SMBH.

\section{Comparison of Hydrodynamic and Gravitational Wave Torques}

The angular momentum of a binary system consisting of a SMBH and a
compact star is given by $J=\mu\sqrt{GMa}$, where $M=M_{\rm
SMBH}+M_c\approx M_{\rm SMBH}$ is the total mass, $\mu=M_{\rm
SMBH}M_c/M\approx M_c$ is the reduced mass, and $a$ is the radius of
the orbit.  The angular frequency of the orbit is equal to
$\sqrt{GM/a^3}$.

The time scale on which $J$ changes as a result of gravitational wave
emission is (e.g. Shapiro \& Teukolsky 1983)
$$
t_{gw}={J\over|dJ/dt|}={5\over32}{c^5\over G^3}{a^4\over M^2\mu}.
\eqno (11)
$$
Let us express this in practical units.  We write $a=10r_1R_g$, and
define $P_2=P/100$ s, where $P=\pi/\Omega$ is the period of the
quadrupolar gravitational waves (this period is equal to one half the
period of the orbit).  Then
$$
t_{gw}=24{m_6^2r_1^4\over m_1}\,{\rm yr}=0.35{P_2^{8/3}\over
m_1m_6^{2/3}}\,{\rm yr}. \eqno (12)
$$
The time to merger is given by
$$
t_m={t_{gw}\over8}=3.0{m_6^2r_1^4\over m_1}\,{\rm
yr}=0.044{P_2^{8/3}\over m_1m_6^{2/3}}\,{\rm yr}. \eqno (13)
$$

LISA or a similar gravitational wave detector is likely to be able to
follow a merger event for at most a few years.  Let us assume $t_m<10$
yr.  The time scale on which the gravitational wave torque acts on the
orbit is given by $t_{gw}=8t_m<100$ yr.  The time scale of the
hydrodynamic torque (eq 10) is seen to be longer than $t_{gw}$ by
nearly ten orders of magnitude (recall that $\dot m\lsim0.1$ for an
ADAF and is probably $\sim10^{-2} -10^{-4}$ in normal galactic nuclei).
Thus it is clear that the hyrdrodynamic torque is completely
negligible.

For a more quantitative consideration, let us compute the number of
wave periods prior to the merger of the two objects:
$$
n_m=\int_0^{t_m} {dt\over P} = {8\over5}{t_m\over P}. \eqno (14)
$$
Putting in numerical values,
$$
n_m = 3.1\times 10^5{m_6r_1^{5/2}\over m_1}
    = 2.2\times10^4{P_2^{5/3}\over m_1m_6^{2/3}}
    = 1.6\times10^5{t_{m,yr}^{5/8}\over m_1^{3/8}m_6^{1/4}}, \eqno (15)
$$
where $t_{m,yr}=t_m/1$ yr.  With a strong signal and a theoretically
computed template of the wave train one expects to follow the phase of
the gravitational waves with a time resolution of better than a
period.  Optimistically, one might be able to resolve down to a tenth
of a wave period (Kip Thorne, private communication).  The importance
of the hydrodynamic drag is then estimated by the quantity
$$
\epsilon\equiv10n_m{t_{gw}\over t_{hd}}=2.8\times10^{-5} {\dot m
m_1^{5/8}t_{m,yr}^{13/8}\over m_6^{5/4}}. \eqno (16)
$$
If $\epsilon>1$, we expect the hydrodynamic perturbation to have a
noticeable effect on the inspiral wave form.  For any reasonable
choice of the parameters, however, we see that $\epsilon\ll1$,
implying that the hydrodynamic drag has a negligible effect on the
observed orbital decay of the compact star.

\section{Discussion}

In this paper we have obtained a very robust result, namely that a
compact star orbiting inside an advection-dominated accretion flow
around a supermassive black hole experiences a negligible amount of
frictional drag from hydrodynamic forces.  In deriving this result we
made conservative assumptions (e.g. we did not take into account the
mitigating effects mentioned at the end of \S2) and, where possible,
we erred on the side of overestimating the drag (e.g. in our estimate
of $I$ in eq 7).  Yet we found that the perturbation due to
hydrodynamic forces, as measured by the parameter $\epsilon$ (eq 16),
is always extremely small.  To determine how relevant this result is
for gravitational wave experiments, we need to consider several
questions.

First, what fraction of galactic nuclei contain SMBHs?  Recent work
has revealed that the fraction is close to unity.  In our local group
of galaxies, the three largest galaxies, namely M31, M32 and our own
Milky Way Galaxy, have dark masses in their nuclei with masses in the
range $10^{6.5}-10^{7.5} M_\odot$.  Outside the local group,
observations with the Hubble Space Telescope and with ground-based
telescopes have revealed dark massive objects in the nuclei of nearly
all galaxies that are accessible to sensitive observations
(e.g. Magorrian et al. 1998, Richstone et al. 1998); the masses are in
the range $10^6-10^{9.5}M_\odot$.  It is considered highly likely that
all these dark masses are SMBHs, though this is yet to be proved
conclusively.  In the case of our own Galactic nucleus and the nucleus
of NGC 4258, the argument for a SMBH is quite compelling (Genzel et
al. 1997, Ghez et al. 1998, Miyoshi et al. 1995, Narayan et al. 1998).

Second, what fraction of galactic nuclei are inactive or only weakly
active, thus indicating highly sub-Eddington accretion?  (We consider
a nucleus to be very sub-Eddington if its luminosity is less than a
few percent of the Eddington luminosity, $L_{Edd}=
10^{46}(M_{SMBH}/10^8M_\odot)~{\rm erg\,s^{-1}}$; when the mass of the
SMBH is not known, we take the luminosity limit to be $10^{43}~{\rm
erg\,s^{-1}}$, which corresponds to 3\% of $L_{Edd}$ for a
$10^{6.5}M_\odot$ SMBH.)  The fraction of inactive/weakly active
nuclei appears to be close to unity.  For instance, there are no
active galactic nuclei in our local group of galaxies or in the nearby
Virgo cluster of galaxies.  (M87, the dominant galaxy in Virgo, has
what may be considered an active nucleus, but its luminosity is highly
sub-Eddington for a $2\times10^9M_\odot$ SMBH, cf. Reynolds et
al. 1996.)  The number density of quasars averaged over a larger
volume of the nearby universe is also extremely low (Krolik 1999).  If
we include the more numerous Seyferts, which have lower luminosities
than quasars but still fall under the definition of active nuclei, the
local number density is of order $2-3\%$ of the number density of all
galaxies brighter than $L_*$ (Huchra \& Burg 1992).  The incidence of
nuclear activity increases with increasing redshift, reaching a peak
at $z\sim2-3$.  However, out to $z\sim1$, the redshift range
accessible to LISA for stellar inspiral events, the fraction of
galaxies with active nuclei is very likely no more than $0.1-0.2$.
Thus, over the volume of the universe of interest to us, the majority
of galactic nuclei have significantly sub-Eddington accretion.

Third, how strong is the evidence that SMBHs with highly sub-Eddington
accretion have ADAFs around them?  In the opinion of this author, the
evidence is fairly strong.  For low mass accretion rates such as we
are considering, $\dot M<0.1-0.01\dot M_{\rm Edd}$, only two stable,
self-consistent, rotating, accretion flow solutions are known (Chen et
al. 1995): the thin disk solution and the ADAF solution.  For all low
luminosity black holes for which sufficient data are available,
whether they be in galactic nuclei or in stellar-mass binaries, some
version of the ADAF model appears to explain the observations
(Narayan, Yi \& Mahadevan 1995, Narayan, McClintock \& Yi 1996,
Reynolds et al. 1996, Narayan, Barret \& McClintock 1997, Di Matteo \&
Fabian 1997, Manmoto et al. 1997, Hameury et al. 1998, Narayan et
al. 1998, Di Matteo et al. 1999, Quataert et al. 1999).  In several
objects, the evidence suggests that both an ADAF and a thin disk are
present, but such that the ADAF is located close to the black hole,
while the thin disk is restricted to radii beyond tens, hundreds or
even thousands of $R_g$. For the gravitational wave experiments that
are the focus of this paper, the compact stars will be in orbits with
radii smaller than about $10R_g$.  This region of the accretion flow
is not in the form of a thin disk in any low luminosity black hole
that has been studied so far; the evidence suggests that the accretion
flow forms an ADAF in all cases.

Finally, if the accretion is not in the form of an ADAF close to the
SMBH, what other form might it take, and could the hydrodynamic drag
be significant?  This is a difficult question to answer since at
present the only solutions we are aware of for low values of $\dot m$
are the thin disk and the ADAF.  One other formal solution is known,
due to Bondi (1952), but it corresponds to the unlikely case of zero
angular momentum in the accreting gas.  In any case, the hydrodynamic
drag force on an orbiting star from a Bondi flow is even less (by a
factor of $\alpha$) than that from an ADAF.

A point worth making is that if the accretion occurs in a non-ADAF
mode, then the mass accretion rate must be a lot less than the values
$\dot M_{\rm SMBH}\sim10^{-2}-10^{-4}\dot M_{\rm Edd}$ that we have
assumed in this paper.  This is because the {\it luminosities} of the
``normal '' nuclei we are considering are extraordinarily low.  For
instance, the SMBH in the nucleus of our own Galaxy has a luminosity
$< 10^{-7}L_{\rm Edd}$ (e.g. Narayan et al. 1998) despite an estimated
mass accretion rate of $\sim 10^{-4}\dot M_{\rm Edd}$ (Quataert,
Narayan \& Reid 1999, and references therein).  The mismatch between
the luminosity and the accretion rate is natural with an ADAF since
the accretion flow advects the missing energy through the event
horizon into the black hole.  Indeed, this argument has been used to
argue for the presence of an event horizon in this and other black
hole systems with ADAFs (see Menou, Quataert \& Narayan 1999 for a
review).  With a non-ADAF model, the accretion rate would have to be
$< 10^{-7}\dot M_{\rm Edd}$ in our Galactic nucleus, and comparably
small in most other low luminosity nuclei.  Even without knowing the
details of the flow, it is probably safe to assume that the
hydrodynamic drag on an orbiting star in such an ultra-low-$\dot M$
accretion flow would be insignificant.

We thus conclude that the majority of inspiral events that LISA might
detect will be unaffected by hydrodynamic interactions.

\noindent{\it Acknowledgments.}  It is a pleasure to thank Kip Thorne
for instigating this study, for useful discussions, and for persuading
the author to publish the results, John Huchra for advice on the
statistics of active galactic nuclei, and Eve Ostriker for help in
understanding the hydrodynamic drag force.  This work was supported in
part by grant AST 9820686 from the NSF.

\bigskip\bigskip
{
\footnotesize
\StartRef
\noindent {\large \bf References} \\

\Ref Abramovici, A., et al. 1992, Science, 256, 325 \\ 

\Ref Abramowicz, M., Chen, X., Kato, S., Lasota, J. P., \& Regev, O.,
1995, ApJ, 438, L37 \\

\Ref Abramowicz, M., Czerny, B., Lasota, J. P., \& Szuszkiewicz,
E. 1988, ApJ, 332, 646 \\

\Ref Bender, P., et al. 1996, LISA. Laser Interferometer Space Antenna
for the detection and observation of gravitational waves, Max-Planck
Institut fur Quantenoptik, Report No. MPQ 208, Garching, Germany \\

\Ref Blandford, R. D., \& Begelman, M. C., 1999, MNRAS, 303, L1 \\ 

\Ref Bondi, H. 1952, MNRAS, 112, 195 \\

\Ref Chakrabarti, S. K. 1993, ApJ, 411, 610 \\

\Ref Chakrabarti, S. K. 1996, Phys. Rev. D53, 2901 \\

\Ref Chen, X., Abramowicz, M., \& Lasota, J. P. 1997, ApJ, 476, 61 \\

\Ref Chen, X., Abramowicz, M., Lasota, J. P., Narayan, R., \& Yi,
I. 1995, ApJ, 443, L61 \\

\Ref Cutler, C. 1998, Phys. Rev. D57, 7089 \\ 

\Ref Danzmann, K., et al. 1996, LISA Pre-Phase A Report,
Max-Planck Institut fur Quantenoptik, Report No. MPQ 208, Garching,
Germany \\ 

\Ref Di Matteo, T., \& Fabian, A. C. 1997, MNRAS, 286, L50 \\

\Ref Di Matteo, T., Quataert, E., Allen, S. W., Narayan, R., \& Fabian,
A. C., 1999, ApJ, submitted (astro-ph/9905052) \\

\Ref Esin, A. A., McClintock, J. E., \& Narayan, R. 1997, ApJ, 489,
867 \\ 

\Ref Folkner, W. M., ed. 1998, Laser-Interferometer Space
Antenna, AIP Conf Proc. 456 (New York: AIP Press) \\ 

\Ref Genzel, R., Eckart, A., Ott, T., \& Eisenhauer, F. 1997, MNRAS,
291, 219 \\

\Ref Ghez, A. M., Klein, B. L., Morris, M., \& Becklin, E. E. 1998,
ApJ, 509, 678 \\

\Ref Hameury, J. M., Lasota, J. P., McClintock, J. E., \& Narayan,
R. 1997, ApJ, 489, 234 \\

\Ref Hils, D., \& Bender, P. 1995, ApJL, 445, L7 \\ 

\Ref Huchra, J., \& Burg, R. 1992, ApJ, 393, 90 \\

\Ref Ichimaru, S. 1977, ApJ, 214, 840 \\

\Ref Kato, S., Fukue, J., \& Mineshige, S. 1998, Black-Hole Accretion
Disks (Kyoto: Kyoto Univ. Press) \\

\Ref Krolik, J. 1999, Active Galactic Nuclei (Princeton: Princeton Univ.) \\

\Ref Magorrian, J., et al. 1998, AJ, 115, 2285 \\

\Ref Manmoto, T., Mineshige, S., \& Kusunose, M. 1997, ApJ, 489, 791 \\

\Ref Menou, K., Quataert, E., \& Narayan, R. 1999, in Black Holes,
Gravitational Radiation, and the Universe, eds. B.R. Iyer \& B. Bhawal
(Dordrecht: Kluwer) \\

\Ref Miyoshi, M., Moran, J., Herrnstein, J., Greenhill, L., Nakai, N.,
Diamond, P., \& Inoue, M. 1995, Nature, 373, 127 \\

\Ref Molteni, D., Gerardi, G., \& Chakrabarti, S. K. 1994, ApJ, 436,
249 \\

\Ref Narayan, R., Barret, D., \& McClintock, J. E. 1997, ApJ, 482, 448 \\ 

\Ref Narayan, R., Kato, S. \& Honma, F. 1997, ApJ, 476, 49 \\

\Ref Narayan, R., Mahadevan, R., Grindlay, J. E., Popham, R. G., \&
Gammie, C. 1998a, ApJ, 492, 554 \\ 

\Ref Narayan, R., Mahadevan, R., \& Quataert, E. 1998, in Theory of
Black Hole Accretion Disks, eds. M. A. Abramowicz, G. Bjornsson, \&
J. E. Pringle, p148 (Cambridge Univ. Press) \\

\Ref Narayan, R., McClintock, J. E., \& Yi, I. 1996, ApJ, 457, 821 \\

\Ref Narayan, R., \& Yi, I. 1994, ApJ, 428, L13 \\ 

\Ref Narayan, R., \& Yi, I. 1995a, ApJ, 444, 231 \\

\Ref Narayan, R., \& Yi, I. 1995b, ApJ, 452, 710 \\ 

\Ref Narayan, R., Yi, I., \& Mahadevan, R. 1995, Nature, 374, 623 \\ 

\Ref Novikov, I. D., \& Thorne, K. S. 1973, in Blackholes, eds. C. DeWitt,
\& B. DeWitt, p343 (Gordon \& Breach) \\

\Ref Ostriker, E. C. 1999, ApJ, 513, 252 \\

\Ref Quataert, E., Di Matteo, T., Ho, L., \& Narayan, R. 1999, ApJ,
submitted \\

\Ref Quataert, E., \& Narayan, R. 1999, ApJ, in press
(astro-ph/9810136) \\

\Ref Quataert, E., Narayan, R., \& Reid, M. J. 1999, ApJ, 517, L101 \\

\Ref Rees, M. J., Phinney, E. S., Begelman, M. C., \& Blandford,
R. D. 1982, Nature, 295, 17 \\

\Ref Reynolds, C. S., Di Matteo, T., Fabian, A. C., Hwang, U., \&
Canizares, C. R. 1996, MNRAS, 283, L111 \\ 

\Ref Richstone, D., et al. 1998, Nature, 395A, 14 \\

\Ref Ryan, F. D. 1995, Phys. Rev. D52, 5707 \\

\Ref Shakura, N. I., \& Sunyaev, R. A. 1973, A\&A, 24, 337 \\

\Ref Shapiro, S. L., \& Teukolsky, S. A. 1983, Black Holes, White
Dwarfs, and Neutron Stars (New York: Wiley) \\

\Ref Sigurdsson, S. 1997, Class. Q. Grav., 14, 1425 \\

\end{document}